\documentclass[10pt]{iopartmod}
\usepackage{amssymb}
\usepackage{amsmath}
\usepackage{makeidx}
\usepackage[dvips]{graphicx}

\input{tcilatex}

\begin{document}
\title{Resonance enhancement of magnetic Faraday rotation}
\author{A. Figotin and I. Vitebskiy}

\begin{abstract}
Magnetic Faraday rotation is widely used in optics. In natural transparent
materials, this effect is very weak. One way to enhance it is to incorporate
the magnetic material into a periodic layered structure displaying a high-Q
resonance. One problem with such magneto-optical resonators is that a
significant enhancement of Faraday rotation is inevitably accompanied by
strong ellipticity of the transmitted light. More importantly, along with the
Faraday rotation, the resonator also enhances linear birefringence and
absorption associated with the magnetic material. The latter side effect can
put severe limitations on the device performance. From this perspective, we
carry out a comparative analysis of optical microcavity and a slow wave
resonator. We show that slow wave resonator has a fundamental advantage when
it comes to Faraday rotation enhancement in lossy magnetic materials.

\end{abstract}
\maketitle

\section{Introduction}

Nonreciprocal effects play a crucial role in microwave technology and optics.
They are absolutely essential in numerous devices such as isolators,
circulators, phase shifters, etc. Nonreciprocal effects in optics and MW occur
in magnetically ordered materials or in the presence of magnetic field
\cite{LLEM,Gurev}. A well-known example is magnetic Faraday rotation related
to nonreciprocal circular birefringence. At optical frequencies, all
nonreciprocal effects are usually weak, and can be further obscured by
absorption, linear birefringence, etc. A\ way to enhance a weak Faraday
rotation is to place the magnetic material in an optical resonator \cite{MPC
Inoue06,Lyubch,MPC Inoue99,Levy07,Levy06,Grishin07,Grishin04,Vinogr}. An
intuitive explanation for the resonance enlacement invokes a simple idea that
in a high-Q optical resonator filled with magnetic material, each individual
photon resides much longer compared to the same piece of magnetic material
taken out of the resonator. Since the nonreciprocal circular birefringence is
independent of the direction of light propagation, one can assume that the
magnitude of Faraday rotation is proportional to the photon residence time in
the magnetic material. With certain reservations, the above assumption does
provide a hand-waving explanation of the phenomenon of resonance enhancement
of weak Faraday rotation. The problem, though, is that the above qualitative
picture of the phenomenon does not apply to composite structures in which the
size of magnetic components is comparable to or lesser than the
electromagnetic wavelength. At the same time, all the existing methods of
resonance enhancement of nonreciprocal effects involve some kind of composite
dielectric structures, such as periodic layered arrays, in which the magnetic
layer thickness is comparable to the electromagnetic wavelength.

In this paper we analyze and compare two qualitatively different approaches to
resonance enhancement of magnetic Faraday rotation. The first one is based on
a magnetic microcavity sandwiched between a pair of Bragg reflectors, as shown
in Fig. \ref{MC}. The second approach is based on a slow wave resonance in a
magnetic photonic crystal, an example of which is shown in Fig. \ref{AFStack}.
Either approach involves spatially periodic composite structures, usually,
periodic layered arrays. In either case, the thickness of magnetic layers
producing Faraday rotation has to be comparable to the electromagnetic
wavelength. Otherwise, the resonance enhancement becomes inefficient. On the
other hand, a linearly polarized electromagnetic wave entering such a magnetic
composite structure inevitably acquires ellipticity, which in this case is a
manifestation of nonreciprocity \cite{PRB08}. If the resonance Q-factor is
high enough, the acquired ellipticity becomes so significant that the very
term "Faraday rotation" becomes irrelevant. Indeed, one cannot assign a
meaningful rotation angle to a wave with nearly circular polarization. The
above circumstance, though, does not diminish the practical importance of the
nonreciprocal effect, which now reduces to the conversion of linear
polarization of the incident wave to nearly circular polarization of
transmitted and/or reflected waves.

The resonance conditions can result in enhancement of the nonreciprocal
effect, which is a desirable outcome. On the other hand, the same resonance
conditions will also cause enhancement of absorption and linear birefringence
in the same magnetic material, which is undesirable. Linear birefringence, if
present, can drastically suppress the Faraday rotation, or any other
manifestation of nonreciprocal circular birefringence. Even more damaging can
be absorption. In uniform magnetic materials, the absorption contributes to
the ellipticity of transmitted wave and causes circular dichroism. In low-loss
magnetic materials those effects can be insignificant. Under the resonance
condition, though, the role of absorption can change dramatically. Firstly,
the enhanced absorption reduces the intensity of light transmitted through the
optical resonator. Secondly, even moderate absorption can lower the Q-factor
of the resonance by several orders of magnitude and, thereby, significantly
compromise its performance as Faraday rotation enhancer. The role of
absorption essentially depends on whether it is a microcavity or a photonic
slow wave resonator. It turns out that the resonance cavity shown in Fig.
\ref{MC} is particularly vulnerable to absorption. This and related questions
are also addressed in this paper.

\section{Notations, definitions, and physical assumptions}

\subsection{Transverse electromagnetic waves in stratified media}

Our analysis is based on the time-harmonic Maxwell equations%
\begin{equation}
\nabla\times\vec{E}\left(  \vec{r}\right)  =i\frac{\omega}{c}\hat{\mu}\left(
\vec{r}\right)  \vec{H}\left(  \vec{r}\right)  ,\;\nabla\times\vec{H}\left(
\vec{r}\right)  =-i\frac{\omega}{c}\hat{\varepsilon}\left(  \vec{r}\right)
\vec{E}\left(  \vec{r}\right)  , \label{THME}%
\end{equation}
where the second rank tensors $\hat{\varepsilon}\left(  \vec{r}\right)  $ and
$\hat{\mu}\left(  \vec{r}\right)  $ are coordinate dependent. In a stratified
medium
\[
\hat{\varepsilon}\left(  \vec{r}\right)  =\hat{\varepsilon}\left(  z\right)
,\hat{\mu}\left(  \vec{r}\right)  =\hat{\mu}\left(  z\right)  ,
\]
where the Cartesian coordinate $z$ is normal to the layers. We also assume
that the dielectric permittivity and magnetic permeability tensors in each
layer has the following form%
\begin{equation}
\hat{\varepsilon}=\left[
\begin{array}
[c]{ccc}%
\varepsilon_{xx} & \varepsilon_{xy} & 0\\
\varepsilon_{yx} & \varepsilon_{yy} & 0\\
0 & 0 & \varepsilon_{zz}%
\end{array}
\right]  ,~~\hat{\mu}=\left[
\begin{array}
[c]{ccc}%
\mu_{xx} & \mu_{xy} & 0\\
\mu_{yx} & \mu_{yy} & 0\\
0 & 0 & \mu_{zz}%
\end{array}
\right]  , \label{eps mu trnsv}%
\end{equation}
in which case the layered structure support transverse electromagnetic waves
with%
\begin{equation}
\vec{E}\left(  \vec{r}\right)  =\vec{E}\left(  z\right)  \perp z,\ \vec
{H}\left(  \vec{r}\right)  =\vec{H}\left(  z\right)  \perp z, \label{EH trsv}%
\end{equation}
propagating along the $z$ direction. The Maxwell equations (\ref{THME}) in
this case reduce to the following system of four ordinary differential
equations%
\begin{equation}
\frac{\partial}{\partial z}\Psi\left(  z\right)  =i\frac{\omega}{c}M\left(
z\right)  \Psi\left(  z\right)  , \label{ME4}%
\end{equation}
\ where%
\begin{equation}
\Psi\left(  z\right)  =\left[
\begin{array}
[c]{c}%
E_{x}\left(  z\right) \\
E_{y}\left(  z\right) \\
H_{x}\left(  z\right) \\
H_{y}\left(  z\right)
\end{array}
\right]  , \label{Psi}%
\end{equation}
and%
\begin{equation}
M\left(  z\right)  =\left[
\begin{array}
[c]{cccc}%
0 & 0 & \mu_{xy}^{\ast} & \mu_{yy}\\
0 & 0 & -\mu_{xx} & -\mu_{xy}\\
-\varepsilon_{xy}^{\ast} & -\varepsilon_{yy} & 0 & 0\\
\varepsilon_{xx} & \varepsilon_{xy} & 0 & 0
\end{array}
\right]  . \label{M trsv}%
\end{equation}
The $4\times4$ matrix $M\left(  z\right)  $ is referred to as the (reduced)
Maxwell operator.

Solutions for the reduced time-harmonic Maxwell equation (\ref{ME4}) can be
presented in the following form%
\begin{equation}
\Psi\left(  z\right)  =T\left(  z,z_{0}\right)  \Psi\left(  z_{0}\right)  ,
\label{T(zz0)}%
\end{equation}
where the $4\times4$ matrix $T\left(  z,z_{0}\right)  $ is the \emph{transfer
matrix}. The transfer matrix (\ref{T(zz0)}) uniquely relates the values of
electromagnetic field (\ref{Psi}) at any two points $z$ and $z_{0}$ of the
stratified medium.

In a uniform medium, the Maxwell operator $M$ in (\ref{M trsv}) is independent
of $z$. In this case, the transfer matrix $T\left(  z,z_{0}\right)  $ can be
explicitly expressed in terms of the respective Maxwell operator $M$%
\begin{equation}
T\left(  z,z_{0}\right)  =\exp\left[  i\frac{\omega}{c}\left(  z-z_{0}\right)
M\right]  . \label{T hmg}%
\end{equation}
In particular, the transfer matrix of an individual uniform layer $m$ is%
\begin{equation}
T_{m}=\exp\left(  i\frac{\omega}{c}z_{m}M_{m}\right)  , \label{Tm}%
\end{equation}
where $z_{m}$ is the thickness of the $m$-th layer.

The transfer matrix $T_{S}$ of an arbitrary stack of layers is a sequential
product of the transfer matrices $T_{m}$ of the constituent layers%
\begin{equation}
T_{S}=\prod_{m}T_{m}. \label{TS}%
\end{equation}

In the following subsection we specify the form of the material tensors
(\ref{eps mu trnsv}), which determine the transfer matrices of the individual
layers and the entire periodic structure. In this paper, we use the same
notations as in our previous publication \cite{PRE01,PRB03,PRB08} related to
magnetic layered structures.

\subsection{Permittivity and permeability tensors of the layers}

We assume that the permittivity and permeability tensors of individual layers
have the following form%
\begin{equation}
\hat{\varepsilon}=\left[
\begin{array}
[c]{ccc}%
\varepsilon+\delta & i\alpha & 0\\
-i\alpha & \varepsilon-\delta & 0\\
0 & 0 & \varepsilon_{zz}%
\end{array}
\right]  ,~~\hat{\mu}=1, \label{eps AF, mu AF}%
\end{equation}
where $\alpha$ is responsible for nonreciprocal circular birefringence and
$\delta$ describes linear birefringence. In a lossless medium, the physical
quantities $\varepsilon$, $\alpha$, and $\delta$ are real. If the direction of
magnetization is changed for the opposite, the parameters $\alpha$ also
changes its sign and so will the sense of Faraday rotation \cite{LLEM,Gurev}.
The absorption, is accounted for by allowing $\varepsilon$, $\alpha$, and
$\delta$ to be complex.

Substitution of (\ref{eps AF, mu AF}) into (\ref{M trsv}) yields the following
expression for the Maxwell operator%
\begin{equation}
M=\left[
\begin{array}
[c]{cccc}%
0 & 0 & 0 & 1\\
0 & 0 & -1 & 0\\
i\alpha & -\varepsilon+\delta & 0 & 0\\
\varepsilon+\delta & i\alpha & 0 & 0
\end{array}
\right]  . \label{M AF}%
\end{equation}
The respective four eigenvectors are
\begin{equation}
\left\{
\begin{array}
[c]{c}%
1\\
-ir_{1}\\
in_{1}r_{1}\\
n_{1}%
\end{array}
\right\}  \leftrightarrow n_{1},\ \left\{
\begin{array}
[c]{c}%
1\\
-ir_{1}\\
-in_{1}r_{1}\\
-n_{1}%
\end{array}
\right\}  \leftrightarrow-n_{1},\ \left\{
\begin{array}
[c]{c}%
-ir_{2}\\
1\\
-n_{2}\\
-in_{2}r_{2}%
\end{array}
\right\}  \leftrightarrow n_{2},\ \left\{
\begin{array}
[c]{c}%
-ir_{2}\\
1\\
n_{2}\\
in_{2}r_{2}%
\end{array}
\right\}  \leftrightarrow-n_{2}. \label{EV AF}%
\end{equation}
$\allowbreak\allowbreak$where%
\begin{equation}
n_{1}=\sqrt{\varepsilon+\sqrt{\delta^{2}+\alpha^{2}}},\ \ n_{2}=\sqrt
{\varepsilon-\sqrt{\delta^{2}+\alpha^{2}}}, \label{n_1, n_2}%
\end{equation}%
\begin{equation}
r_{1}=\frac{\alpha}{\sqrt{\delta^{2}+\alpha^{2}}+\delta},\ r_{2}=\frac
{\sqrt{\delta^{2}+\alpha^{2}}-\delta}{\alpha}, \label{r_1, r_2}%
\end{equation}
Compared to \cite{Levy07}, we use slightly different notations.

The explicit expression for the transfer matrix $\hat{T}(A)$ of a single
uniform layer of thickness $A$ is%
\begin{equation}
\hat{T}(A)=\hat{W}\left(  A\right)  \hat{W}^{-1}(0), \label{TG}%
\end{equation}
where%
\begin{equation}
\hat{W}\left(  A\right)  =\left[
\begin{array}
[c]{llll}%
e^{i\phi_{1}} & e^{-i\phi_{1}} & -ir_{2}e^{i\phi_{2}} & -ir_{2}e^{-i\phi_{2}%
}\\
-ir_{1}e^{i\phi_{1}} & -ir_{1}e^{-i\phi_{1}} & e^{i\phi_{2}} & e^{-i\phi_{2}%
}\\
ir_{1}n_{1}e^{i\phi_{1}} & -ir_{1}n_{1}e^{-i\phi_{1}} & -n_{2}e^{i\phi_{2}} &
n_{2}e^{-i\phi_{2}}\\
n_{1}e^{i\phi_{1}} & -n_{1}e^{-i\phi_{1}} & -ir_{2}n_{2}e^{i\phi_{2}} &
ir_{2}n_{2}e^{-i\phi_{2}}%
\end{array}
\right]  , \label{WG}%
\end{equation}
and%
\[
\phi_{1}=\frac{\omega}{c}An_{1},\ \ \phi_{2}=\frac{\omega}{c}An_{2}.
\]

The eigenvectors (\ref{EV AF}) correspond to elliptically polarized states.
There are two important particular cases corresponding to linearly and
circularly polarized eigenmodes, respectively.

\subsubsection{Non-magnetic medium with linear birefringence}

In the case of a non-magnetic medium%
\begin{equation}
\alpha=0,\ r_{1}=0,\ r_{2}=0. \label{alpha=0}%
\end{equation}
The respective eigenmodes are linearly polarized%
\begin{equation}
\left\{
\begin{array}
[c]{c}%
1\\
0\\
0\\
n_{1}%
\end{array}
\right\}  \leftrightarrow n_{1},\ \left\{
\begin{array}
[c]{c}%
1\\
0\\
0\\
-n_{1}%
\end{array}
\right\}  \leftrightarrow-n_{1},\ \left\{
\begin{array}
[c]{c}%
0\\
1\\
-n_{2}\\
0
\end{array}
\right\}  \leftrightarrow n_{2},\ \left\{
\begin{array}
[c]{c}%
0\\
1\\
n_{2}\\
0
\end{array}
\right\}  \leftrightarrow-n_{2}. \label{EV A0}%
\end{equation}
where%
\[
n_{1}=\sqrt{\varepsilon+\delta},\ \ n_{2}=\sqrt{\varepsilon-\delta}.
\]

\subsubsection{Magnetic medium with circular birefringence}

Another important limiting case corresponds to a uniaxial magnetic medium with%
\begin{equation}
\delta=0,\ r_{1}=1,\ r_{2}=1. \label{delta=0}%
\end{equation}
The respective eigenmodes are circularly polarized
\begin{equation}
\left\{
\begin{array}
[c]{c}%
1\\
-i\\
in_{1}\\
n_{1}%
\end{array}
\right\}  \leftrightarrow n_{1},\ \left\{
\begin{array}
[c]{c}%
1\\
-i\\
-in_{1}\\
-n_{1}%
\end{array}
\right\}  \leftrightarrow-n_{1},\ \left\{
\begin{array}
[c]{c}%
-i\\
1\\
-n_{2}\\
-in_{2}%
\end{array}
\right\}  \leftrightarrow n_{2},\ \left\{
\begin{array}
[c]{c}%
-i\\
1\\
n_{2}\\
in_{2}%
\end{array}
\right\}  \leftrightarrow-n_{2}. \label{EV F}%
\end{equation}
where%
\[
n_{1}=\sqrt{\varepsilon+\alpha},\ \ n_{2}=\sqrt{\varepsilon-\alpha}.
\]

\subsection{Numerical values of material tensors}

Our objectives include two distinct problems associated with Faraday rotation enhancement.

One problem can be caused by the presence of linear birefringence described by
the parameter $\delta$ in (\ref{eps AF, mu AF}). Linear birefringence $\delta$
competes with circular birefringence $\alpha$. At optical frequencies, the
former can easily prevail and virtually annihilate any manifestations of
nonreciprocal circular birefringence. If linear birefringence occurs in
magnetic F layers in Fig. \ref{AFStack}, it can be offset by linear
birefringence in the alternating dielectric A layers. Similarly, in the case
of a magnetic resonance cavity in Fig. \ref{MC}, the destructive effect of the
linear birefringence in the magnetic D layer can be offset by linear
birefringence in layers constituting the Bragg reflectors. In either case, the
cancellation of linear birefringence of the magnetic layers only takes place
at one particular frequency. Therefore, the layered structure should be
designed so that this particular frequency coincides with the operational
resonance frequency of the composite structure. The detailed discussion on the
effect of linear birefringence and ways to deal with it will be presented elsewhere.

In the rest of the paper we will focus on the problem associated with
absorption. This problem is unrelated to the presence or absence of linear
birefringence and, therefore, can be handled separately. For this reason, in
our numerical simulation we can set $\delta=0$ and use the following
expressions for the dielectric permittivity tenors of the magnetic F-layers
and dielectric A-layers in Fig. \ref{AFStack}
\begin{equation}
\hat{\varepsilon}_{F}=\left[
\begin{array}
[c]{ccc}%
\varepsilon_{F}+i\gamma & i\alpha & 0\\
-i\alpha & \varepsilon_{F}+i\gamma & 0\\
0 & 0 & \varepsilon_{3}%
\end{array}
\right]  ,\label{eps_F}%
\end{equation}%
\begin{equation}
\hat{\varepsilon}_{A}=\left[
\begin{array}
[c]{ccc}%
\varepsilon_{A} & 0 & 0\\
0 & \varepsilon_{A} & 0\\
0 & 0 & \varepsilon_{A}%
\end{array}
\right]  ,\label{eps_A}%
\end{equation}
where $\varepsilon_{F}$, $\varepsilon_{A}$, and $\gamma$ are real. Parameter
$\gamma$ describes absorption of the magnetic material.

In the case of photonic cavity in Fig. \ref{MC} we use similar material
parameters. The permittivity tensor of the magnetic D-layer is the same as
that of the magnetic F-layers in Fig. \ref{AFStack}%
\begin{equation}
\hat{\varepsilon}_{D}=\hat{\varepsilon}_{F}\label{eps_D}%
\end{equation}
$\hat{\varepsilon}_{F}$ is defined in (\ref{eps_F}). The permittivity tensors
of the alternating dielectric layers A and B constituting the Bragg reflectors
in Fig. \ref{MC} are chosen as follows%
\begin{equation}
\hat{\varepsilon}_{B}=\left[
\begin{array}
[c]{ccc}%
\varepsilon_{B} & o & 0\\
0 & \varepsilon_{B} & 0\\
0 & 0 & \varepsilon_{B}%
\end{array}
\right]  ,\ \ \hat{\varepsilon}_{C}=\left[
\begin{array}
[c]{ccc}%
\varepsilon_{C} & 0 & 0\\
0 & \varepsilon_{C} & 0\\
0 & 0 & \varepsilon_{C}%
\end{array}
\right]  .\label{eps_A, eps_B}%
\end{equation}

In either case, only the magnetic layers F or D are responsible for
absorption, which is a realistic assumption.

In the case of periodic stack in Fig. \ref{AFStack} we use the following
numerical values of the diagonal components of the permittivity tensors%
\begin{equation}
\varepsilon_{F}=5.37,\ \ \varepsilon_{A}=2.1.\label{Num PS}%
\end{equation}
Similar values are used in the case of photonic microcavity in Fig. \ref{MC}%
\[
\varepsilon_{D}=\varepsilon_{C}=5.37,\ \ \varepsilon_{B}=2.1.
\]
The numerical values of the gyrotropic parameter $\alpha$, as well as the
absorption coefficient $\gamma$ of the magnetic layers F and D, remain
variable. We also tried different layer thicknesses $d_{A}$,$\ d_{F}$%
,$\ d_{B}$, $d_{C}$, and $\ d_{D}$. But in this paper we only include the
results corresponding to the following numerical values%
\begin{equation}
d_{A}=d_{C}=0.8L,\ d_{F}=d_{C}=0.2L,\ d_{D}=0.4L,\label{Num d}%
\end{equation}
where $L$ is the length of a unit cell of the periodic array%
\[
L=d_{F}+d_{A}=d_{B}+d_{C}.
\]
The thickness $d_{D}$ of the defect layer in Fig. \ref{MC} is chosen so that
the frequency of the defect mode falls in the middle of the lowest photonic
band gap of Bragg reflectors.

\subsection{Scattering problem for magnetic layered structure}

In all cases, the incident wave $\Psi_{I}$ propagates along the $z$ direction
normal to the layers. Unless otherwise explicitly stated, the incident wave
polarization is linear with $\vec{E}_{I}\parallel x$. Due to the nonreciprocal
circular birefringence of the magnetic material, the transmitted and reflected
waves $\Psi_{P}$ and $\Psi_{R}$ will be elliptically polarized with the
ellipse axes being at an angle with the $x$ direction.

The transmitted and reflected waves, as well the electromagnetic field
distribution inside the layered structure, are found using the transfer matrix
approach. Let us assume that the left-hand and the right-hand boundaries of a
layered array are located at $z=0$ and $a=d$, respectively. According to
(\ref{T(zz0)}) and (\ref{TS}), the incident, transmitted, and reflected waves
are related as follows%
\begin{equation}
\Psi_{P}(d)=T_{S}\left(  \Psi_{I}(0)+\Psi_{R}(0)\right)  . \label{BC}%
\end{equation}
Knowing the incident wave $\Psi_{I}$ and the transfer matrix $T_{S}$ of the
entire layered structure and assuming, we can solve the system (\ref{BC}) of
four linear equations and, thereby, find the reflected and transmitted waves.
Similarly, using the relation (\ref{T(zz0)}), we can also find the field
distribution inside the layered structure.

The transmission and reflection coefficients of the slab (either uniform, or
layered) are defined as follows%
\begin{equation}
t=\frac{S_{P}}{S_{I}},\ r=-\frac{S_{R}}{S_{I}}, \label{t,r}%
\end{equation}
where $S_{I}$, $S_{P}$, and $S_{R}$ are the Poynting vectors of the incident,
transmitted, and reflected waves, respectively. The slab absorption is
\begin{equation}
a=1-t-r. \label{a}%
\end{equation}

If the incident wave polarization is linear, the coefficients $t$, $r$, and
$a$ are independent of the orientation of vector $\vec{E}_{I}$ in the $x-y$
plane, because for now, we neglect the linear birefringence $\delta$. Due to
nonreciprocal circular birefringence, the polarization of the transmitted and
reflected waves will always be elliptic.

By contrast, if the incident wave polarization is circular, the coefficients
$t$, $r$, and $a$ depend on the sense of circular polarization. The
polarization of the transmitted and reflected waves in this case will be
circular with the same sense of rotation as that of the incident wave.

The effect of nonreciprocal circular birefringence on transmitted wave can be
quantified by the following expression%
\begin{equation}
\Delta\Psi_{P}=\frac{1}{2}\left[  \left(  \Psi_{P}\right)  _{\alpha}-\left(
\Psi_{P}\right)  _{-\alpha}\right]  \label{NR Diff}%
\end{equation}
where $\left(  \Psi_{P}\right)  _{\alpha}$ and $\left(  \Psi_{P}\right)
_{-\alpha}$ respectively correspond to the wave transmitted through the
original periodic structure and through the same structure but with the
opposite sign of circular birefringence parameter $\alpha$. If the incident
wave polarization is linear with $\vec{E}_{I}\parallel x$, the vector-column
(\ref{NR Diff}) has the following simple structure%
\[
\Delta\Psi_{P}=\left(  \vec{E}_{P}\right)  _{y}\left[
\begin{array}
[c]{c}%
0\\
1\\
1\\
0
\end{array}
\right]  ,
\]
implying that the $y$ component $\left(  \vec{E}_{P}\right)  _{y}$ of the
transmitted wave has "purely" nonreciprocal origin and, therefore, can used to
characterize the magnitude of nonreciprocal circular birefringence on
transmitted wave. Indeed, in the absence of magnetism, the parameter $\alpha$
in (\ref{eps AF, mu AF}), (\ref{eps_F}), and \ref{eps_D}) vanishes and the
transmitted wave is linearly polarized with $\vec{E}_{P}\parallel x$. The
above statement follows directly from symmetry consideration and remains valid
even in the presence of linear birefringence $\delta$ in (\ref{eps AF, mu AF}%
). Further in this paper will use the ratio%
\begin{equation}
\rho=\frac{\left(  E_{P}\right)  _{y}}{\left(  E_{I}\right)  _{x}},\text{
\ where \ }\left\vert \rho\right\vert <1. \label{rho}%
\end{equation}
to characterize the effect circular birefringence on transmitted wave.

Generally, the transmitted wave polarization in the situation in Figs.
\ref{AFStack} and \ref{MC} is elliptical, rather than linear. Therefore, the
quantity $\rho$ in (\ref{rho}) is not literally the sine of the Faraday
rotation angle. Let us elaborate on this point. The electromagnetic eigenmodes
of the layered structures in Figs. \ref{AFStack} and \ref{MC} with
permittivity tensors given in (\ref{eps_F}) through (\ref{eps_A, eps_B}) are
all circularly polarized. This implies that if the polarization of the
incident wave is circular, the transmitted and reflected waves will also be
circularly polarized. On the other hand, due to the nonreciprocal (magnetic)
effects, the transmission/reflection coefficients for the right-hand circular
polarization are different from those for the left-hand circular polarization.
This is true regardless of the presence or absence of absorption. Consider now
\ a linearly polarized incident wave. It can be viewed as a superposition of
two circularly polarized waves with equal amplitudes. Since the
transmission/reflection coefficients for the right-hand and left-hand circular
polarizations are different, the transmitted and reflected waves will be
elliptically polarized. Such an ellipticity develops both in the case of a
uniform slab and in the case of a layered stack, periodic or aperiodic, with
or without absorption. Note, though, that at optical frequencies, the dominant
contribution to ellipticity of the wave transmitted through a uniform slab is
usually determined by absorption, which is largely responsible for circular
dichroism. Without absorption, the ellipticity of the wave transmitted through
a uniform magnetic slab would be negligible. This might not be the case for
the layered structures in Figs. \ref{AFStack} and \ref{MC} at frequencies of
the respective transmission resonances. In these cases, the ellipticity of
transmitted and reflected waves can be significant even in the absence of
absorption. Moreover, if the Q-factor of the respective resonance is high
enough, the transmitted wave polarization becomes very close to circular and,
therefore, cannot be assigned any meaningful angle of rotation. The numerical
examples of the next section illustrate the above statements.

To avoid confusion, note that a linear polarized wave propagating in a
uniform, lossless, unbounded, magnetic medium (\ref{eps_F}) will not develop
any ellipticity. Instead, it will display a pure Faraday rotation. But the
slab boundaries and the layer interfaces will produce some ellipticity even in
the case of lossless magnetic material. The absorption provides an additional
contribution to the ellipticity of transmitted and reflected waves. The latter
contribution is referred to as circular dichroism.

For simplicity, in further consideration we will often refer to the quantity
$\rho$ in (\ref{rho}) as the amount of (nonreciprocal) Faraday rotation,
although, due to the ellipticity, it is not exactly the sine of the Faraday
rotation angle.

In all plots, the frequency $\omega$ and the Bloch wave number $k$ are
expressed in dimensionless units of $cL^{-1}$ and $L^{-1}$, respectively. In
our computations we use a transfer matrix approach identical to that described
in Ref. \cite{PRE01,PRB03}.

\section{Resonance enhancement of circular birefringence}

\subsection{Cavity resonance: Lossless case}

Let us start with the resonance enhancement based on microcavity. The magnetic
layer D in Fig. \ref{MC} is sandwiched between two identical periodic stacks
playing the role of distributed Bragg reflectors. The D-layer is also referred
to as a defect layer, because without it, the layered structure in Fig.
\ref{MC} would be perfectly periodic. The thickness of the defect layer is
chosen so that the microcavity develops a single resonance mode with the
frequency lying in the middle of the lowest photonic band gap of the adjacent
periodic stacks. This resonance mode is nearly localized in the vicinity of
the magnetic D-layer.

A typical transmission spectrum of such a layered structure in the absence of
absorption is shown in Fig. \ref{t_FSDM_A0_G}. The stack transmission develops
a sharp peak at the defect mode frequency. The respective transmission
resonance is accompanied by a dramatic increase in field amplitude in the
vicinity of the magnetic D-layer. The large field amplitude implies the
enhancement of magnetic Faraday rotation produced by the D-layer, as clearly
seen in Fig. \ref{p_FSDM_A_G0}.

If the Q-factor of the microcavity exceeds certain value and/or if the
circular birefringence of the magnetic material of the D-layer is strong
enough, the resonance frequency of the defect mode splits into two, as shown
in Fig. \ref{p_FSDM_A_G0}(c) and (d). Each of the two resonances is associated
with left or right circular polarization. The transmitted light will also
display nearly perfect circular polarization with the opposite sense of
rotation for the twin resonances. Formally, the above nonreciprocal effect
cannot be classified as Faraday rotation, but it does not diminish its
practical value.

\subsection{Slow wave resonance: Lossless case}

The second approach to Faraday rotation enhancement is based on the
transmission band edge resonance in periodic stacks of magnetic layers
alternating with some other dielectric layers, as shown in Fig. \ref{AFStack}.
A typical transmission spectrum of such a layered structure is shown in Fig.
\ref{t_PFS_a0_g}. The sharp peaks in transmission bands correspond to
transmission band edge resonances, also known as Fabry-Perot resonances. The
resonance frequencies are located close to a photonic band edge, where the
group velocity of the respective Bloch eigenmodes is very low. This is why the
transmission band edge resonances are referred to as slow wave resonances. All
resonance frequencies are located in transmission bands -- not in photonic
band gaps, as in the case of a localized defect mode. The resonance field
distribution inside the periodic stack is close to a standing wave composed of
a pair of Bloch modes with equal and opposite group velocities and nearly
equal large amplitudes%
\begin{equation}
\Psi_{T}\left(  z\right)  =\Psi_{k}\left(  z\right)  +\Psi_{-k}\left(
z\right)  ,\label{1sw}%
\end{equation}
The left-hand and right-hand photonic crystal boundaries coincide with the
standing wave nodes, where the forward and backward Bloch components interfere
destructively to meet the boundary conditions. The most powerful slow wave
resonance corresponds to the transmission peak closest to the respective
photonic band edge, where the wave group velocity is lowest. At resonance, the
energy density distribution inside the periodic structure is typical of a
standing wave%
\begin{equation}
W\left(  z\right)  \propto W_{I}N^{2}\sin^{2}\left(  \frac{\pi}{NL}z\right)
,\label{W(z) RBE}%
\end{equation}
where $W_{I}$ is the intensity of the incident light, $N$ is the total number
of unit cells (double layers) in the periodic stack in Fig. \ref{AFStack}.

Similarly to the case of magnetic cavity resonance, the large field amplitude
implies the enhancement of magnetic Faraday rotation produced by magnetic
F-layers, as demonstrated in Fig. \ref{p_PFS_a_g0}. Again, if the Q-factor of
the slow wave resonance exceeds certain value and/or if the circular
birefringence $\alpha$ of the magnetic material of the F-layers is strong
enough, each resonance frequency splits into two, as shown in Fig.
\ref{t_PFS_a_g0_fwLx}. Each of the two twin resonances is associated with left
or right circular polarization. To demonstrate it, let us compare the
transmission dispersion in Fig. \ref{t_PFS_a_g0_fwLx}, where the incident
light polarization is linear, to the transmission dispersion in Figs.
\ref{t_PFS_a_g0_fwLcp} and \ref{t_PFS_a_g0_fwLcn}, where the incident wave is
circularly polarized. One can see that the case in Fig. \ref{t_PFS_a_g0_fwLx}
of linearly polarized incident light reduces to a superposition of the cases
in Figs. \ref{t_PFS_a_g0_fwLcp} and \ref{t_PFS_a_g0_fwLcn} of two circularly
polarized incident waves with opposite sense of rotation.

\subsection{The role of absorption}

In the absence of absorption, the practical difference between cavity
resonance and slow wave resonance is not that obvious. But if the magnetic
material displays an appreciable absorption, the difference between the two
resonators becomes huge. The physical reason for this is as follows.

In the case of a slow wave resonance, the reduction of the transmitted wave
energy is mainly associated with absorption. Indeed, although some fraction of
the incident light energy is reflected at the left-hand interface of the
periodic stack in Fig. \ref{AFStack}, this fraction remains limited even in
the case of strong absorption, as seen in Fig. \ref{ar_PFS_a0_g}(b). So, the
main source of the energy losses in a slow wave resonator is absorption, which
is a natural side effect of the Faraday rotation enhancement (some important
reservations can be found in \cite{PRB08}).

\begin{figure}[tbph]
\scalebox{0.8}{\includegraphics[viewport=-50 0 500 300,clip]{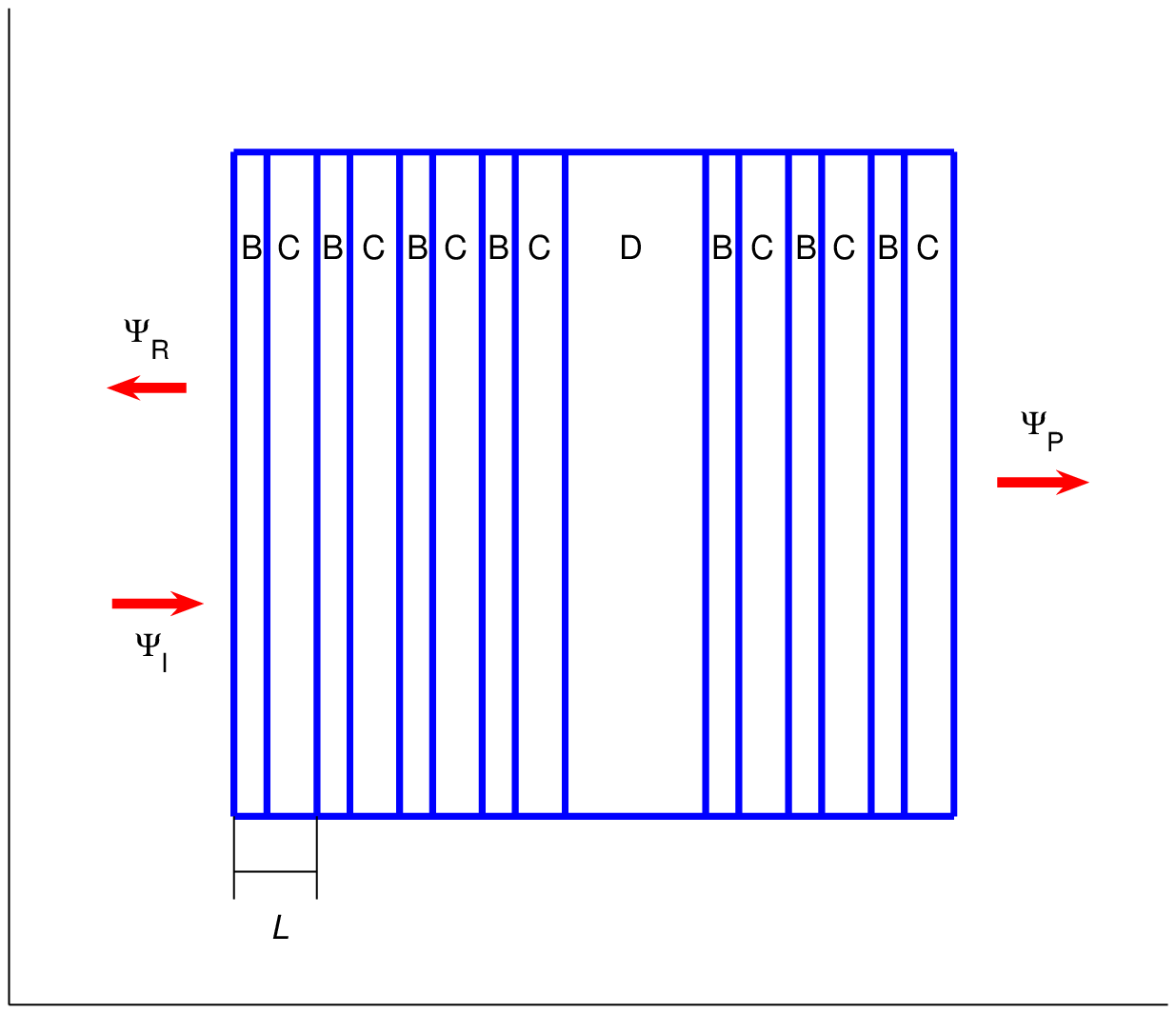}}
\caption{(Color online) Magnetic resonance cavity composed of magnetic layer D
sandwiched between a pair of identical periodic non-magnetic stacks (Bragg
reflectors). The incident wave $\Psi_{I}$ is linearly polarized with
$E\parallel x$. Due to the nonreciprocal circular birefringence of the
magnetic material of D-layer, the reflected wave $\Psi_{R}$ and the
transmitted wave $\Psi_{P}$ are both elliptically polarized.}
\label{MC}
\end{figure}

\begin{figure}[tbph]
\scalebox{0.8}{\includegraphics[viewport=-50 0 500 300,clip]{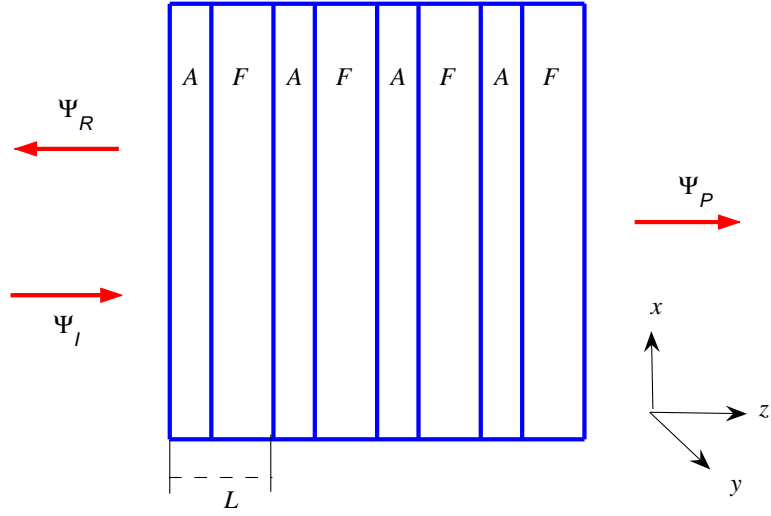}}
\caption{(Color online) Periodic layered structure composed of alternate
magnetic (F) and dielectric (A) layers. The F-layers are made of the same
lossy magnetic material as the D-layer in Fig. \ref{MC}. $L$ is the unit cell
length. The incident wave $\Psi_{I}$ is linearly polarized with $E\parallel
x$. Due to the nonreciprocal circular birefringence of the magnetic material
of the F-layers, the reflected wave $\Psi_{R}$ and the transmitted wave
$\Psi_{P}$ are both elliptically polarized.}
\label{AFStack}
\end{figure}

\begin{figure}[tbph]
\scalebox{0.8}{\includegraphics[viewport=-50 0 500 300,clip]{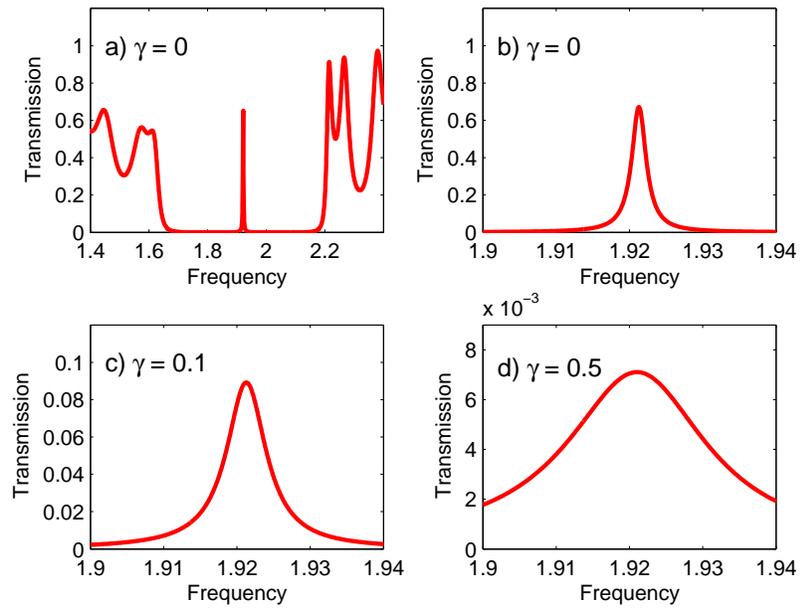}}
\caption{(Color online) Transmission dispersion of the layered array in Fig.
\ref{MC} for different values of absorption coefficient $\gamma$ of the
D-layer. Circular birefringence $\alpha$ is negligible. Fig. (b) shows the
enlarged portion of Fig. (a) covering the vicinity of microcavity resonance.}
\label{t_FSDM_A0_G}
\end{figure}

\begin{figure}[tbph]
\scalebox{0.8}{\includegraphics[viewport=-50 0 500 300,clip]{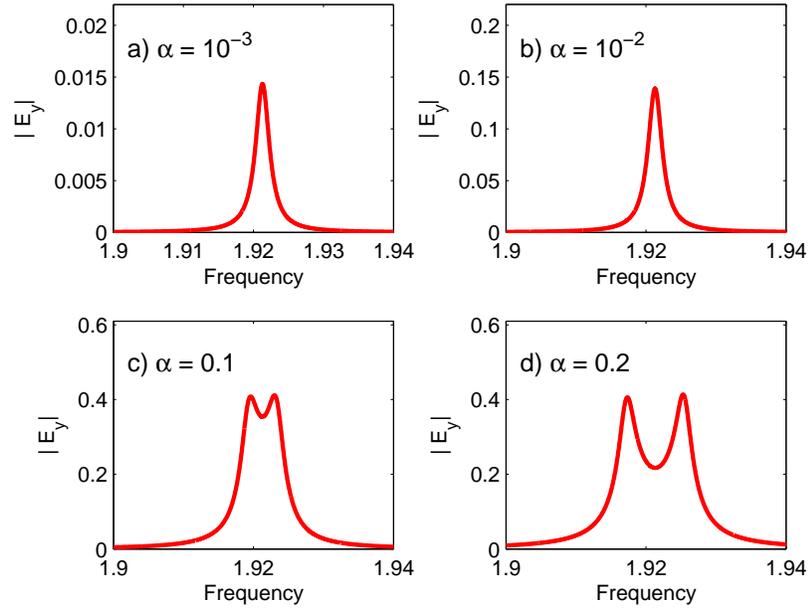}}
\caption{(Color online) Frequency dependence of polarization component
$\left\vert E_{y}\right\vert $ of the wave transmitted through layered array
in Fig. \ref{MC} for different values of circular birefringence $\alpha$ of
the D-layer and zero absorption. When circular birefringence $\alpha$ is
strong enough, the cavity resonance splits into a pair of twin resonances,
corresponding to two circularly polarized modes with opposite sense of
rotation. The incident wave is linearly polarized with $\vec{E}\parallel x$.}
\label{p_FSDM_A_G0}
\end{figure}

\begin{figure}[tbph]
\scalebox{0.8}{\includegraphics[viewport=-50 0 500 300,clip]{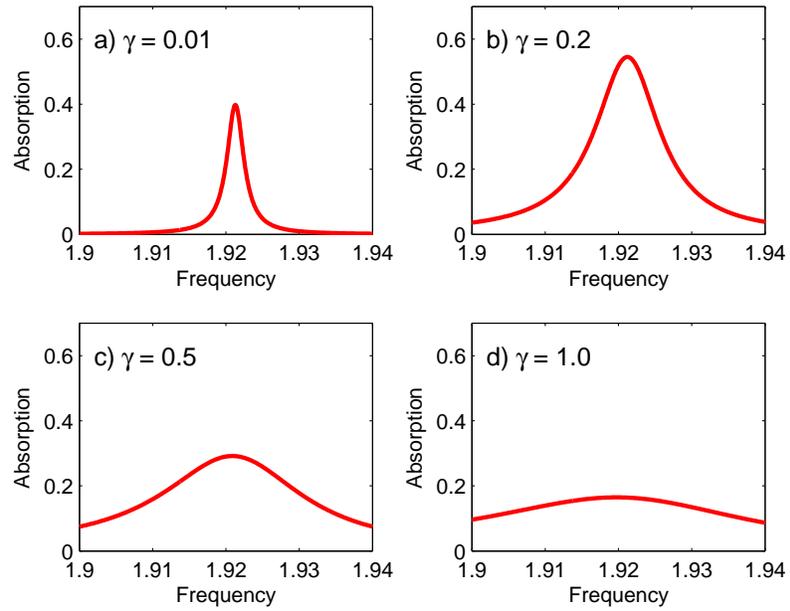}}
\caption{(Color online) Frequency dependence of absorption of the layered
array in Fig. \ref{MC} for different values of absorption coefficient $\gamma$
of the D-layer. Circular birefringence $\alpha$ is negligible. The frequency
range shown covers the vicinity of microcavity resonance. Observe that the
stack absorption decreases after coefficient $\gamma$ exceeds certain value,
which is in sharp contrast with the case of a periodic stack, shown in Figs.
\ref{ar_PFS_a0_g}.}
\label{a_FSDM_A0_G}
\end{figure}

\begin{figure}[tbph]
\scalebox{0.8}{\includegraphics[viewport=-50 0 500 300,clip]{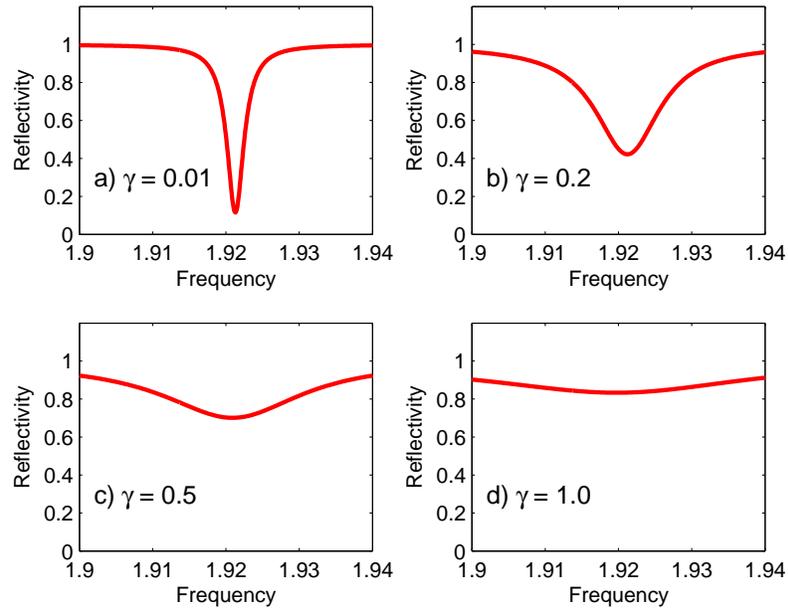}}
\caption{(Color online) Frequency dependence of the reflectance $r$ of the
layered array in Fig. \ref{MC} for different values of absorption coefficient
$\gamma$ of the D-layer. Circular birefringence $\alpha$ is negligible. The
frequency range shown covers the vicinity of microcavity resonance. Observe
that if the absorption coefficient $\gamma$ of D-layer increases, the stack
reflectivity also increases approaching unity. Such a behaivior is line with
frequency dependence of the stack absorption shown in Fig. \ref{a_FSDM_A0_G}.
It is in sharp contrast with the case of a periodic stack, shown in Figs.
\ref{ar_PFS_a0_g}.}
\label{r_FSDM_A0_G}
\end{figure}

\begin{figure}[tbph]
\scalebox{0.8}{\includegraphics[viewport=-50 0 500 300,clip]{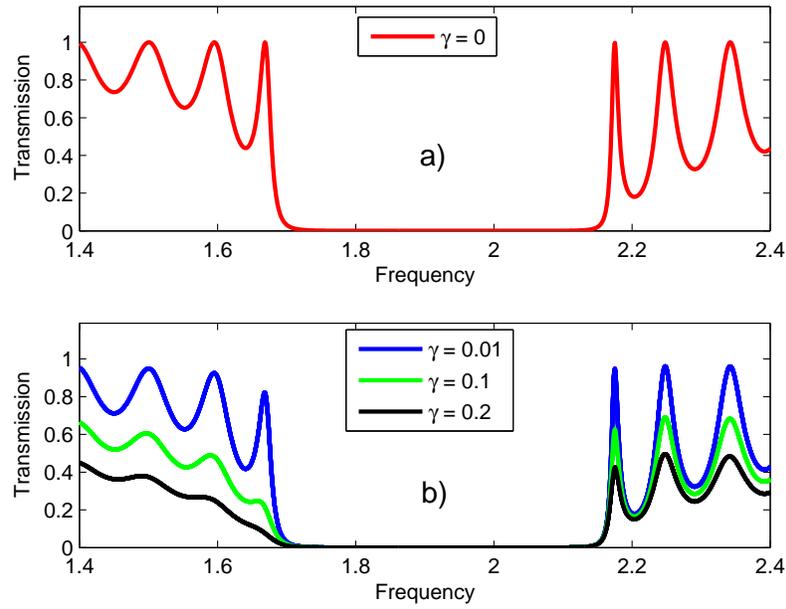}}
\caption{(Color online) Transmission dispersion of periodic layered structure
in Fig. \ref{AFStack} for different values of absorption coefficient $\gamma$
of the F-layers. Circular birefringence $\alpha$ is negligible.}
\label{t_PFS_a0_g}
\end{figure}

\begin{figure}[tbph]
\scalebox{0.8}{\includegraphics[viewport=-50 0 500 300,clip]{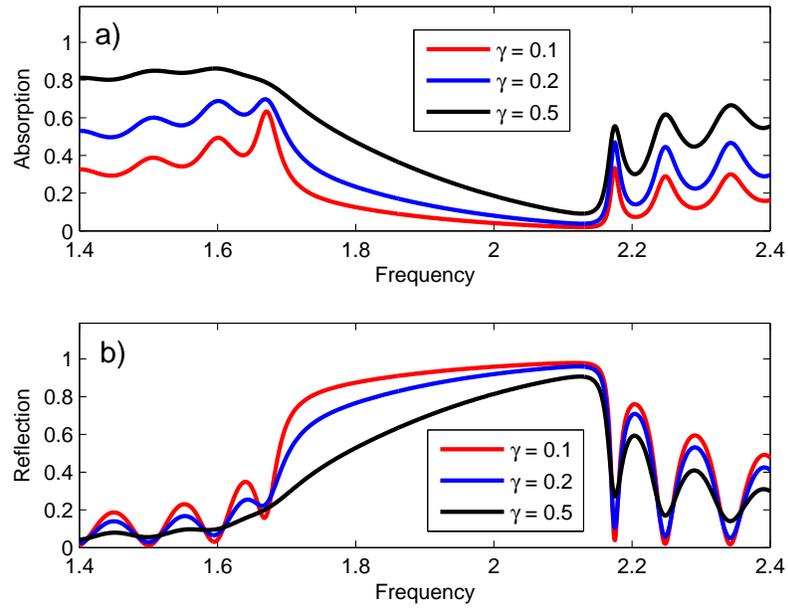}}
\caption{(Color online) Frequency dependence of (a) absorption and (b)
transmission of periodic layered structure in Fig. \ref{AFStack} for different
values of absorption coefficient $\gamma$ of the F-layers. Circular
birefringence $\alpha$ is negligible.}
\label{ar_PFS_a0_g}
\end{figure}

\begin{figure}[tbph]
\scalebox{0.8}{\includegraphics[viewport=-50 0 500 300,clip]{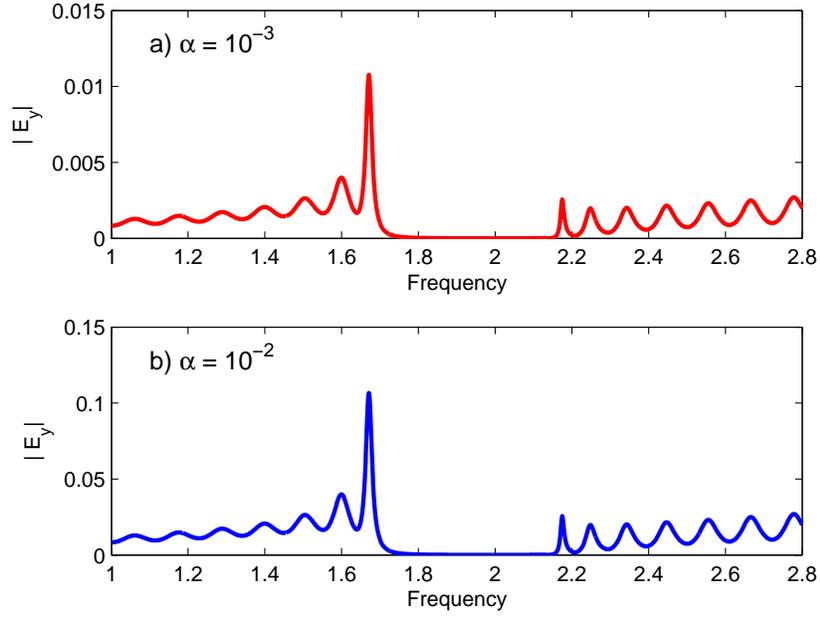}}
\caption{(Color online) Frequency dependence of polarization
component $\left\vert E_{y}\right\vert $ of the wave transmitted through the
periodic layered structure in Fig. \ref{AFStack} for different values of
circular birefringence $\alpha$ of the F-layers and zero absorption. The
incident wave is linearly polarized with $\vec{E}\parallel x$.}
\label{p_PFS_a_g0}
\end{figure}

\begin{figure}[tbph]
\scalebox{0.8}{\includegraphics[viewport=-50 0 500 300,clip]{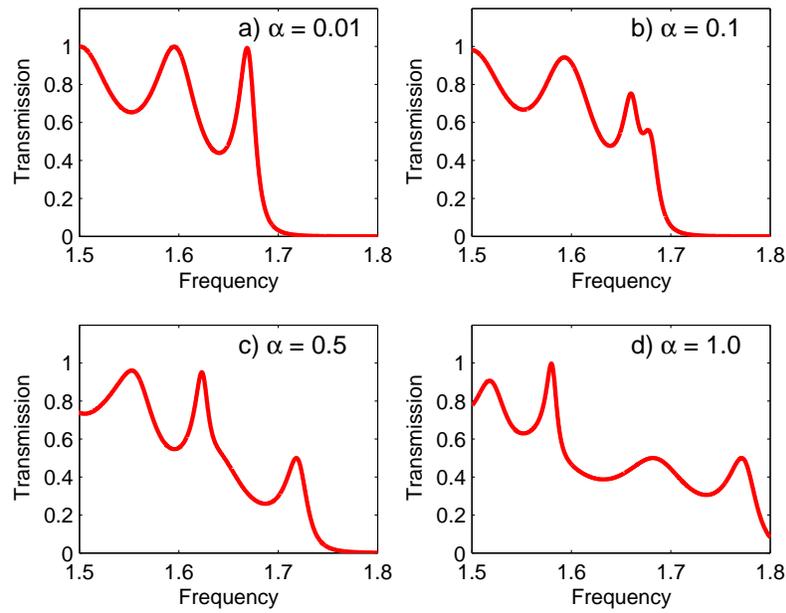}}
\caption{(Color online) Transmission dispersion of periodic layered structure
in Fig. \ref{AFStack} for different values of circular birefringence $\alpha$
of the F-layers and zero absorption. When circular birefringence $\alpha$ is
large enough, each transmission resonance splits into a pair of twin
resonances, corresponding to two circularly polarized modes with opposite
sense of rotation. The incident wave polarization is linear.}
\label{t_PFS_a_g0_fwLx}
\end{figure}

\begin{figure}[tbph]
\scalebox{0.8}{\includegraphics[viewport=-50 0 500 300,clip]{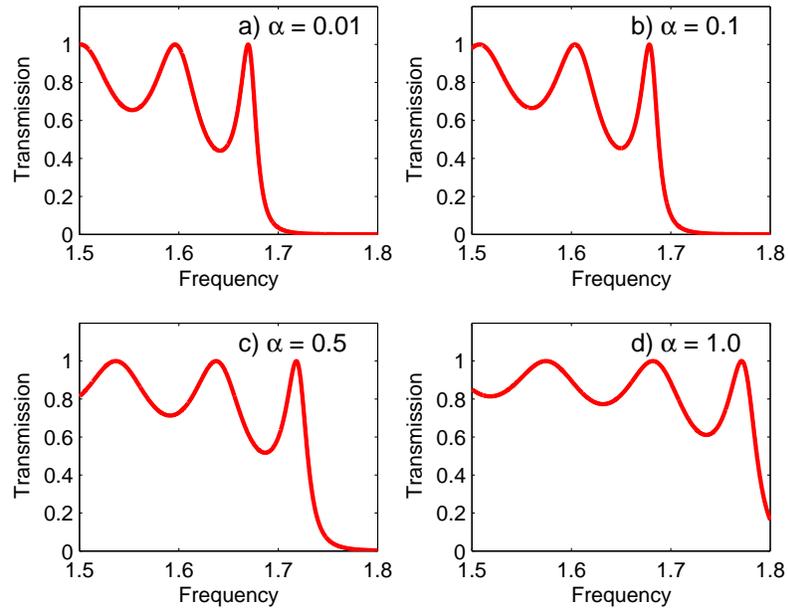}}
\caption{(Color online) The same as in Fig. \ref{t_PFS_a_g0_fwLx}, but the
incident wave polarization is circular with positive sense of rotation.}
\label{t_PFS_a_g0_fwLcp}
\end{figure}

\begin{figure}[tbph]
\scalebox{0.8}{\includegraphics[viewport=-50 0 500 300,clip]{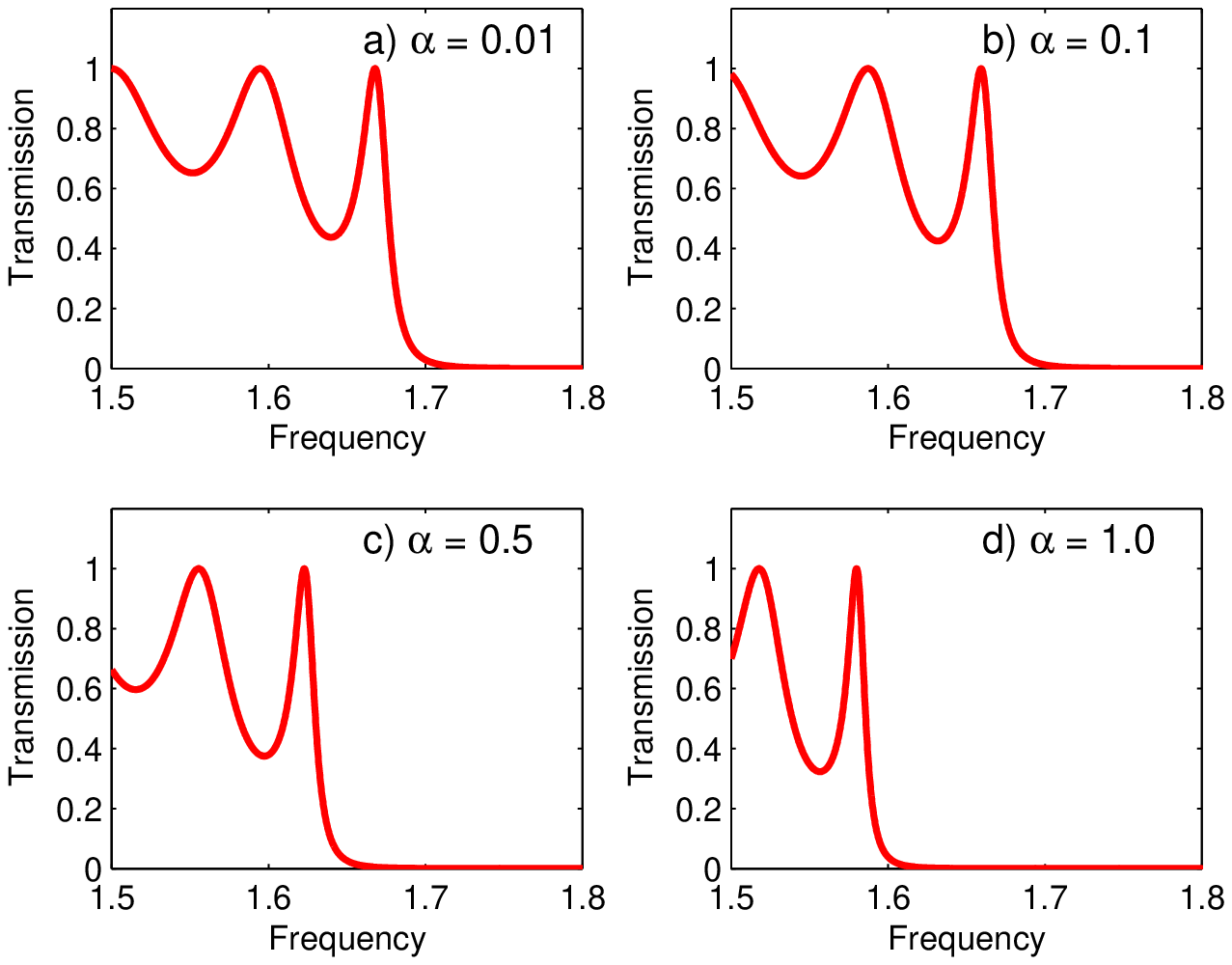}}
\caption{(Color online) The same as in Figs. \ref{t_PFS_a_g0_fwLx} and
\ref{t_PFS_a_g0_fwLcp}, but the incident wave polarization is circular with
negative sense of rotation.}
\label{t_PFS_a_g0_fwLcn}
\end{figure}

In the case of magnetic cavity resonance, the situation is fundamentally
different. In this case, the energy losses associated with absorption cannot
be much different from those of slow wave resonator, provided that both arrays
display comparable enhancement of Faraday rotation. What is fundamentally
different is the reflectivity. An inherent problem with any (localized) defect
mode is that any significant absorption makes it inaccessible. Specifically,
if the D-layer in Fig. \ref{MC} displays an appreciable absorption, the entire
layered array becomes highly reflective. As a consequence, a major portion of
the incident light energy is reflected from the stack surface and never even
reaches the magnetic D-layer. Such a behavior is illustrated in Fig.
\ref{r_FSDM_A0_G}, where we can see that as soon as the absorption coefficient
$\gamma$ exceeds certain value, further increase in $\gamma$ leads to high
reflectivity of the layered structure. In the process, the total absorption
$a$ reduces, as seen in Fig. \ref{a_FSDM_A0_G}, but the reason for this
reduction is that the light simply cannot reach the magnetic layer. There is
no Faraday rotation enhancement in this case.

\textbf{Acknowledgments:} Effort of A. Figotin and I. Vitebskiy is sponsored
by the Air Force Office of Scientific Research, Air Force Materials Command,
USAF, under grant number FA9550-04-1-0359.

\bigskip\pagebreak

\pagebreak%

\pagebreak%

\pagebreak%

\pagebreak%

\pagebreak%

\pagebreak%

\pagebreak%

\pagebreak%

\pagebreak%

\pagebreak%

\pagebreak%

\pagebreak%

\begin{thebibliography}{99}                                                                                               %


\bibitem {LLEM}L. D. Landau, E. M. Lifshitz, L. P. Pitaevskii.
\textsl{Electrodynamics of continuous media}. (Pergamon, N.Y. 1984).

\bibitem {Gurev}A. G. Gurevich and G. A. Melkov. \textsl{Magnetization
Oscillations and Waves}. (CRC Press, N.Y. 1996).

\bibitem {MPC Inoue06}M. Inoue, et all. \textsl{Magnetophotonic crystals
(Topical Review)}. J. Phys. D: Appl. Phys. 39, R151--R161 (2006).

\bibitem {Lyubch}I. Lyubchanskii1, N. Dadoenkova1, M. Lyubchanskii1, E.
Shapovalov, and T. Rasing. \textsl{Magnetic photonic crystals}. J. Phys. D:
Appl. Phys. 36, R277--R287 (2003)

\bibitem {MPC Inoue99}M. Inoue, K. Arai, T. Fuji, and M. Abe.
\textsl{One-dimensional magnetophotonic crystals}. J. Appl. Phys. 85, 5768 (1999).

\bibitem {Levy07}M. Levy and A. A. Jalali. \textsl{Band structure and Bloch
states in birefringent onedimensional magnetophotonic crystals: an analytical
approach}. J. Opt. Soc. Am. B, 24, 1603-1609 (2007).

\bibitem {Levy06}M. Levy and R. Li. \textsl{Polarization rotation enhancement
and scattering mechanisms in waveguide magnetophotonic crystals}. Appl. Phys.
Lett. 89, 121,113 (2006)..

\bibitem {Grishin07}S. Khartsev and A. Grishin. \textsl{High performance
magneto-optical photonic crystals}. J. Appl. Phys. 101, 053,906 (2007).

\bibitem {Grishin04}S. Kahl and A. Grishin. \textsl{Enhanced Faraday rotation
in all-garnet magneto-optical photonic crystal}. Appl. Phys. Lett. 84, 1438 (2004).

\bibitem {Vinogr}S. Erokhin, A. Vinogradov, A. Granovsky, and M. Inoue. Field
Distribution of a Light Wave near a Magnetic Defect in One-Dimensional
Photonic Crystals. Physics of the Solid State, 49, 497 (2007).

\bibitem {}Vol. 49

\bibitem {PRE01}A. Figotin, and I. Vitebsky. \textsl{Nonreciprocal magnetic
photonic crystals}. Phys. Rev. \textbf{E}63, 066609 (2001).

\bibitem {PRB03}A. Figotin, and I. Vitebskiy. \textsl{Electromagnetic
unidirectionality in magnetic photonic crystals}. Phys. Rev. \textbf{B}67,
165210 (2003)

\bibitem {PRB08}A. Figotin and I. Vitebskiy. \textsl{Absorption suppression in
photonic crystals}. Phys. Rev. \textbf{B}77, 104421 (2008)
\end{thebibliography}
\end{document}